\documentclass{mn2e}
\usepackage{graphicx}

\def\go{
\mathrel{\raise.3ex\hbox{$>$}\mkern-14mu\lower0.6ex\hbox{$\sim$}}
}
\def\lo{
\mathrel{\raise.3ex\hbox{$<$}\mkern-14mu\lower0.6ex\hbox{$\sim$}}
}
\def\simeq{
\mathrel{\raise.3ex\hbox{$\sim$}\mkern-14mu\lower0.4ex\hbox{$-$}}
}

\def\etal{{et al.\ }}

\def\msun{{\rm M_{\odot}}}

\begin{document}

\title[{\it XMM-Newton} observations of 3C 273]
{{\it XMM-Newton} observations of 3C 273}
\author[K.L. Page \etal]{K.L. Page$^{1}$, M.J.L. Turner$^{1}$, C. Done$^{2}$, P.T. O'Brien$^{1}$, J.N. Reeves$^{3}$, S. Sembay$^{1}$ 
\and and M. Stuhlinger$^{4}$\\
$^{1}$ X-Ray and Observational Astronomy Group, Department of Physics \& Astronomy,  
University of Leicester, LE1 7RH, UK\\
$^{2}$ Department of Physics, University of Durham, South Road, Durham, DH1 3LE, UK\\
$^{3}$ Laboratory for High Energy Astrophysics, Code 662, NASA Goddard Space Flight Center, Greenbelt, MD 20771, USA \\
$^{4}$ Institut f{\" u}r Astronomie und Astrophysik - Astronomie, Sand 1, 72076 T{\" u}bingen, Germany\\
}

\date{Received / Accepted}

\label{firstpage}

\maketitle

\begin{abstract}

A series of nine {\it XMM-Newton} observations of the radio-loud quasar 3C~273 are presented, concentrating mainly on the soft excess. Although most of the individual observations do not show evidence for iron emission, co-adding them reveals a weak, broad line (EW~$\sim$~56~eV). The soft excess component is found to vary, confirming previous work, and can be well fitted with multiple blackbody components, with temperatures ranging between $\sim$~40 and $\sim$~330~eV, together with a power-law. Alternatively, a Comptonisation model also provides a good fit, with a mean electron temperature of $\sim$~350~eV, although this value is higher when the soft excess is more luminous over the 0.5--10~keV energy band. In the RGS spectrum of 3C~273, a strong detection of the O{\sc vii} He$\alpha$ absorption line at zero redshift is made; this may originate in warm gas in the local intergalactic medium, consistent with the findings of both Fang \etal (2003) and Rasmussen \etal (2003). 

\end{abstract}


\section{Introduction}
\label{sec:intro}

3C~273 was the first object positively identified as a quasar, in
1963: the radio source was identified with a magnitude 13,
star-like object by Hazard, Mackey \& Shimmins (1963) and the redshift
measured by Schmidt (1963) to be z~=~0.158. 
3C~273 was later
found to be a X-ray source, by Bowyer \etal (1970) and Kellogg \etal
(1971), and has been observed with X-ray instruments ever
since. It is a radio-loud quasar, with a jet showing superluminal motion. Although many more quasars
have been discovered since the 1960s, 3C~273 remains one of our nearest
neighbours; this, therefore, makes it a prime object to study, over the
entire range of the electromagnetic spectrum; see Courvoisier (1998) and references therein.
Previous X-ray observations found the high-energy continuum could be fitted by a hard power-law, with a variable photon index, $\Gamma$~$\sim$~1.3--1.6 (Turner \etal 1990; Turner \etal 1991; Williams \etal 1992). Observations by {\it EXOSAT} (Turner \etal 1985) first indicated the existence of a soft excess, at energies $\lo$~1~keV; this was subsequently confirmed by further {\it EXOSAT} observations (Courvoisier \etal 1987; Turner \etal 1990), together with data from {\it Einstein} (Wilkes \& Elvis 1987; Turner \etal 1991), {\it Ginga} (Turner \etal 1990) and {\it ROSAT} (Staubert 1992). {\it Beppo-SAX} (Orr \etal 1998) and {\it ASCA} (Yaqoob \etal 1994; Cappi \etal 1998) have also observed 3C~273, along with {\it EUVE}, which allowed the soft excess to be detected down to $\sim$~0.1~keV (Marshall \etal 1995). The actual form of the soft excess could not previously be determined, due to lack of precision in the low-energy instruments: power-law, blackbody and thermal Bremsstrahlung models produced equally acceptable fits. The EPIC (European Photon Imaging Camera) instruments (Str\"{u}der \etal 2001; Turner \etal 2001) on board {\it XMM-Newton} are improving the situation, however, helping to distinguish between different models much more readily.

The soft excess of 3C~273 has been previously found to vary (Turner \etal 1985; Courvoisier \etal 1987; Turner \etal 1990; Grandi \etal 1992; Leach, McHardy \& Papadakis 1995); {\it Ginga} observations (Saxton \etal 1993) found that its fractional variability was larger than that in the corresponding power-law component.

In this paper, nine {\it XMM} observations of 3C~273 are presented.
In Section~\ref{sec:xmmobs} the {\it XMM} datasets are given and the problem of pile-up discussed. Section~\ref{sec:specanal} covers the spectral analysis, concentrating in particular on the soft excess emission. Sections~\ref{sec:rgs} and \ref{sec:om} cover the RGS (Reflection Grating Spectrometer) and optical data respectively, while Section~\ref{sec:disc} discusses what can be determined from the spectral fits, with the final conclusions given in Section~\ref{sec:conc}. Throughout the paper, H$_{0}$ is taken to be 50~km~s$^{-1}$~Mpc$^{-1}$ and q$_{0}$~=~0.

\section{XMM-Newton Observations}
\label{sec:xmmobs}

3C~273 is one of the {\it XMM-Newton} calibration targets, so is
frequently observed by the instruments. At the time of writing, the quasar has been observed in
revolutions 94, 95, 96, 277, 370, 373, 382, 472, 554 and 563, spanning a time
period of two and half years, from 2000-06-13 to 2003-01-05. These data are a combination of public and proprietary calibration observations. Since the PN  was in Timing mode during revolution 382, this observation was excluded from the analysis; MOS 1 was in Timing mode for four of the nine observatiosn, so only MOS 2 data were used. The observation during revolution 472 is considerably shorter than the others; this, combined with the low luminosity of 3C~273 at this time, leads to larger error bars than measured during the other revolutions. Table~\ref{obs} lists the dates of the observations, together with the exposure times and instrumental set-up. The ODFs (Observation Data Files) were
obtained from the online Science Archive\footnote{http://xmm.vilspa.esa.es/external/xmm\_data\_acc/xsa/index.shtml}; the data were then processed  and the
event-lists filtered using {\sc xmmselect} within the {\sc
sas} (Science Analysis Software) v5.4.1.

\begin{table*}
\begin{center}
\caption{The details of the {\it XMM} observations of 3C~273 analysed in this sample. For revolution 563, the two MOS 2 observations were summed. The `clean' exposure times are given, i.e. those after the periods of high background have been removed. Revolution 472 has about half the exposure of the next shortest observation.} 
\label{obs}
\begin{tabular}{p{1.0truecm}p{1.5truecm}p{1.8truecm}p{1.3truecm}p{1.3truecm}p{1.6truecm}p{1.4truecm}}
\hline
Rev. & Obs. ID & start date & \multicolumn{2}{c}{exposure time (ks)} & \multicolumn{2}{c}{filter}\\
 & & & MOS 2 & PN & MOS 2 & PN\\ 
\hline
94 & 0126700301 & 2000-06-13 & 53.4 & 39.7 & medium & medium\\
95  & 0126700701 & 2000-06-15 & 27.7 & 19.6& medium & medium\\
%& 0126700601 & 2000-06-15 &  & & med & med\\
96 & 0126700801 & 2000-06-17 & 42.6 & 31.8& medium & medium\\
277 & 0136550101 & 2001-06-13 & 41.7/36.6 & 30.1& medium/thin & medium\\
370 & 0112770101 & 2001-12-16 & 4.3 & 3.1& medium & thin\\
373 & 0112770201 & 2001-12-22 & 4.1 &3.0 & medium & thin\\
472& 0112770601 & 2002-07-07 & 2.3 & 1.7& medium & thin\\
554 & 0112770801 & 2002-12-17 & 4.5 & 3.3& medium & thin\\
563 (i)& 0136550501 & 2003-01-05 & 7.9 & 5.7& medium & medium\\
563  (ii) & 0112770701 & 2003-01-05 & 4.6 & 3.3& medium & thin\\
\hline
\end{tabular}
\end{center}
\end{table*}

%The most recent (time-dependent) response matrices were used (e.g., {\bf m21$\_$r7$\_$im$\_$p0$\_$2002-11-07.rmf} for observations using MOS 2 after 2002-11-07), together with an ancillary response function obtained through running {\sc arfgen}.

\begin{figure}
\begin{center}
\includegraphics[clip,width=0.9\linewidth,angle=0]{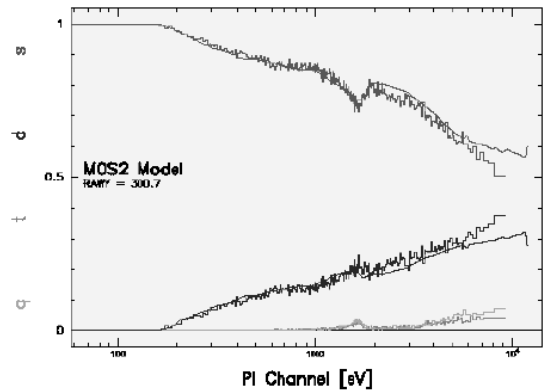}
\caption{This plot shows the result of running {\sc epatplot} on the MOS 2 spectrum of 3C~273 in revolution 96. The top curve shows the expected (solid line) and observed (histogram) fractions of single-pixel events, while the corresponding results for the double events are shown just below. The lowest lines correspond to triple and quadruple events. There are fewer single-events, but more doubles, observed than are predicted; this shows the data are piled-up.}
\label{epatplot}
\end{center}
\end{figure}

Only Small Window Mode observations were used, to minimise the pile-up problems. Upon investigation, it was found that the
MOS spectra were slightly piled-up, even with the quicker read-out time from the Small Window. Figure~\ref{epatplot} shows the
output from the {\sc sas} task {\it epatplot}; the panel
compares the expected fractions of single-, double-, triple- and
quadruple-pixel events (solid line) with those actually measured in
the spectrum (histogram). It can be seen that a smaller than expected
fraction of single events is measured above $\sim$~1~keV, while the reverse is true for doubles; this is
the main signature of pile-up.

Pulse pile-up in CCD cameras occurs when there is a significant probability 
that two or more photons registering within a given CCD frame will have 
overlapping charge distributions. This can lead to a spectral distortion if 
the resulting charge distribution is recognised as a single event whose 
energy is the sum of the overlapping events, or a flux loss if the charge 
distribution has a pattern which is not within the detector's pattern library. The dominant effect of moderate pile-up in the MOS cameras is a loss of flux, with little, or no, spectral distortion, especially if single-pixel events only are analysed.
The degree of pile-up for a given point source depends on the source 
strength and also the point-spread function, throughput, pixel size and frame 
accumulation time of the given X-ray detector system. Approximate limiting 
count rates are quoted in the {\it XMM} User Handbook for the MOS and PN small 
window modes of 5 and 130 counts s$^{-1}$, respectively.

3C 273 has a typical 0.1-10 keV count rate of between 10/35 and 20/65 counts s$^{-1}$ for MOS/PN respectively,  
and is therefore moderately piled-up in the MOS but only very weakly piled-up (if at all) in the PN. Spectral distortion due to pile-up can be reduced by only 
analysing single pixel events, although there is a trade-off in sensitivity.
The spectral fitting has, therefore, been restricted to single pixel events 
in both cameras. The MOS single pixel 
spectra have also been corrected for residual pile-up effects using the method described in Molendi and Sembay (2003). This method uses the fact that the majority of 
diagonal bi-pixel events within a source box are created by the pile-up of 
two single pixel events. The observed diagonal bi-pixel events spectrum can 
then be used to correct the observed single pixel spectrum. Additionally, a development version of the MOS response matrix was used for the observations, which has reduced the systematic errors in the low energy ($<$~2~keV) band; this is expected to be included in the next {\sc sas} release. 

The MOS 2 and PN data were then fitted simultaneously and the results are presented in the following sections. There are well-known systematic differences between MOS and PN fitted spectral parameters, arising from their imperfect calibration and, at this stage, it is not clear which instrument has the smaller systematic errors. The statistical precision of the 3C~273 observations is such that these systematic errors dominate. The practical solution adopted here is to carry out joint fits to MOS and PN, allowing the normalisation to float in order to take care of the differing degrees of flux-loss due to pile-up in the two instruments. Individual fits to the MOS and PN data give better $\chi^2_{\nu}$s, but different parameter values, caused by the calibration uncertainties. For example, over 3--10~keV, there is a slight difference in power-law slopes, with $\Delta \Gamma$~$\leq$~0.1, but using blackbodies to fit the soft excess led to noticeably different temperatures; e.g., $\sim$~95 and 240~eV for PN, compared to $\sim$~130 and 400~eV for MOS. However, the joint fits are acceptable on the basis of $\chi^2$ alone.

\section{Spectral Analysis}
\label{sec:specanal}
\subsection{Iron lines}

The object of this paper is to analyse the soft excess of 3C~273 and to investigate any spectral changes there may be over time. The possible presence of an iron emission line was, however, investigated, by fitting the MOS 2 and PN spectra over the 3--10~keV range with a simple power-law model. On the whole, iron lines, either narrow or broad, were not detected in the individual pointings; the limits on the equivalent widths are given in Table~\ref{ironlines}. An alternative way to analyse the possible changes in the iron emission is to compare the line fluxes for different observations; this separates any variability due to the continuum from the intrinsic line variabilty. The values are also given in Table~\ref{ironlines}, where it can be seen that they are consistent with a constant line flux.

The spectra from all nine observations were then co-added, to search for the possible presence of an iron line; this resulted in a spectrum with a total exposure time of almost 193~ks for the MOS and 127~ks for the PN. (Only those observations taken with the medium filter were used.) As Table~\ref{bb} and Figure~\ref{3-10gamma-lum} show, the 3--10~keV spectral index does not vary greatly between observations ($\Gamma$ lies between 1.64 and 1.78); this is important, since co-adding spectra with different slopes would lead to curvature and, hence, force a broad residual when fitting a simple power-law model. 

\begin{figure}
\begin{center}
\includegraphics[clip,width=0.7\linewidth,angle=-90]{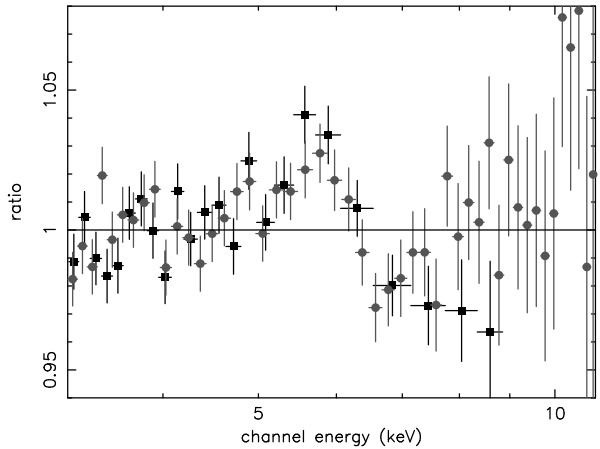}
\caption{A weak broad line can be seen in the co-added spectrum of all nine observations. The MOS data points are shown as squares, the PN as circles.}
\label{coadded}
\end{center}
\end{figure}

\begin{figure}
\begin{center}
\includegraphics[clip,width=0.7\linewidth,angle=-90]{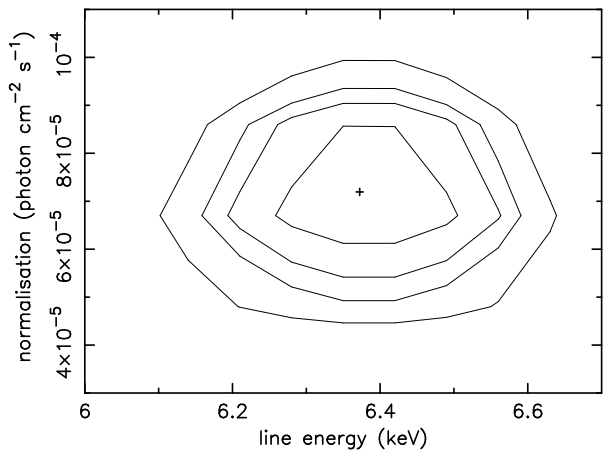}
\caption{A contour plot for the broad line, showing the 68, 90, 95 and 99~per~cent (innermost to outermost curves) possibilities for the energy of the line, for two degrees of freedom. The plot shows that the line is not strongly ionized.}
\label{linecont}
\end{center}
\end{figure}

There is some evidence for a narrow line (E~=~6.34~$\pm$~0.03~keV; EW~=~8~$\pm$~3~eV; $\Delta\chi^{2}$~=~18 for 2 degrees of freedom), but the evidence for a weak broad line was significantly better ($\Delta\chi^{2}$ of 57 for 3 degrees of freedom), giving an equivalent width of 56~$\pm$~19~eV, for E~=~6.37~$\pm$~0.09~keV and $\sigma$~~=~0.60~$\pm$~0.11~keV. Adding an absorption edge (E~=~7.30~$\pm$~0.07~keV; $\tau$~=~0.04~$\pm$~0.01) further imrpoves the fit ($\Delta\chi^{2}$ of 15 for 2 degrees of freedom), giving a final reduced $\chi^{2}$ value of 1451/1728. Figure~\ref{coadded} shows this broad line clearly above a simple power-law model, fitted above 3~keV, while Figure~\ref{linecont} plots the confidence contours for the energy of the line. The implies that the line is not strongly ionised.

Iron emission in 3C~273 was first identified by {\it Ginga} (Turner \etal 1990), where a line of EW~$\sim$~50~eV was detected, although the width could not be constrained. At the time of the {\it Ginga} observation, the 2--10~keV luminosity of the source was measured to be $\sim$~7~$\times$~10$^{45}$ erg~s$^{-1}$, showing 3C~273 to be somewhat fainter than in the present observations (Table~\ref{bb}). Note that the luminosities tabulated here are over the narrower energy band of 3--10~keV.
A weak and broad, but ionised, line was reported by Yaqoob \& Serlemitsos (2000), who detected such a component in {\it RXTE} and {\it ASCA} observations of 3C~273. Kataoka \etal (2002) also confirmed the presence of a broad line in earlier {\it RXTE} data. In the present data, the line appears to be neutral (Figure~\ref{linecont}). The EW of $\sim$~50~eV found in the present data is consistent with these earlier observations. Superluminal blazars, such as Mrk~421, show no hint of iron lines in their X-ray spectra (e.g., Brinkman \etal 2001); these are sources which are very probably being viewed directly down the radio jet. 3C~273 is oriented such that the radio lobes are visible, although superluminal motion is still observed. The presence of the iron line indicates that a Seyfert-like disc spectrum is being seen in 3C~273.

In a recent paper (Page \etal 2003), the decrease in strength of narrow lines with luminosity (i.e., the X-ray Baldwin effect) was discussed. The fact that, in the coadded data presented here, a broad line is preferable to a narrow component is in agreement with this finding. There is also a Baldwin effect for broad lines such as the one measured here (Nandra \etal 1997), which are thought to be produced through reflection off the inner accretion disc; the low EW found for the weak broad line here supports this result.

\begin{table*}
\begin{center}
\caption{The limits on the equivalent widths (EW) of neutral (6.4~keV) and ionised (6.7~keV) iron emission in 3C~273. The narrow lines had an intrinsic width of $\sigma$~=~0.01~keV, the broad lines, 0.5~keV. Where the line was less than 99~per~cent significant, the 90~per~cent upper limit is given. The other error bars are at the 1$\sigma$ level.} 
$^{a}$ flux in units of 10$^{-13}$ erg cm$^{-2}$ s$^{-1}$
\label{ironlines}

\begin{tabular}{p{0.6truecm}p{1.3truecm}p{1.5truecm}p{1.2truecm}p{1.5truecm}p{1.2truecm}p{1.2truecm}}
\hline
Rev.  & \multicolumn{2}{c}{Narrow Line} & \multicolumn{4}{c} {Broad Line}\\
 &   \multicolumn{2}{c}{(neutral)}  & \multicolumn{2}{c}{(neutral)}  & \multicolumn{2}{c}{(ionised)}\\
  & EW (eV) & line flux$^{a}$ & EW (eV) & line flux$^{a}$ & EW (eV) & line flux$^{a}$ \\
\hline
94 & $<$9 & $<$ 0.77 & 47~$\pm$~13 & 3.98~$\pm$~1.49 & $<$51 & $<$4.16\\
95  & $<$10 & $<$ 0.84 &  54~$\pm$~19 & 4.39~$\pm$~1.66 & 42~$\pm$~20 & 3.29~$\pm$~1.73\\ 
96 &  $<$17 & $<$1.42 & 71~$\pm$~20 & 5.84~$\pm$~1.66 & 71~$\pm$~22 & 5.62~$\pm$~1.73\\
277 & $<$12 & $<$1.20  & $<$40 & $<$4.15 & $<$33 & $<$3.32\\
370 &  $<$28 & $<$3.26  & $<$81 & $<$9.37& $<$110 & $<$12.33\\
373 &  $<$41 & $<$4.74  & $<$136 & $<$15.60 & $<$121 & $<$13.49\\
472 & $<$74 & $<$5.83  & $<$166 & $<$13.00 & $<$101& $<$7.80 \\
554 &  $<$36 & $<$3.86 & $<$147 & $<$15.50 & $<$144 & $<$14.74\\ 
563 & $<$21  & $<$1.79 & $<$72 & $<$6.17 & $<$69 & $<$5.75\\
\hline
\end{tabular}
\end{center}
\end{table*}

\subsection{The soft excess}
\label{sec:softexcess}

\begin{figure}
\begin{center}
\includegraphics[clip,width=0.7\linewidth,angle=-90]{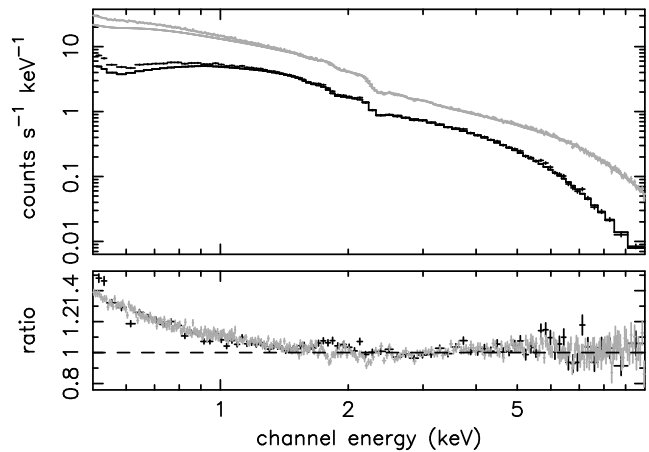}
\caption{An extrapolation of the power-law fitted over the 3--10~keV  band (Table~\ref{bb}) shows the soft excess of 3C~273, during revolution 95. Both MOS 2 (black) and PN (grey) data are shown.}
\label{se}
\end{center}
\end{figure}

As is conventional, after fitting a power-law with Galactic absorption
(N$_{H}$~=~1.79~$\times$~10$^{20}$~cm$^{-2}$; obtained from the {\sc
ftool} {\it nh}, which derives the value from Dickey \& Lockman 1990) to the 3--10~keV energy
band (to avoid the broad soft excess), the fit was extrapolated down to 0.5~keV. (The MOS and PN calibration differences increase rapidly below this point, so lower energy data have not been used.) There is an obvious soft excess, of which Figure~\ref{se}
shows an example.

The soft excess of 3C~273, as observed by EPIC, can be well modelled by multiple blackbody (BB) components. Considering the revolution 95 data as an example, fitting the broad-band spectrum with a power-law together with a single BB component gives an unacceptable fit, with $\chi^{2}$/dof~=~2782/2279. Adding a second BB gives the best-fit value of 2546/2277; the F-test value for this improvement is 18. When using the F-test, a value of F~$>$~3.0 (for one parameter) corresponds to an improvement of $>$~90~per~cent. 

\begin{figure}
\begin{center}
\includegraphics[clip,width=0.7\linewidth,angle=-90]{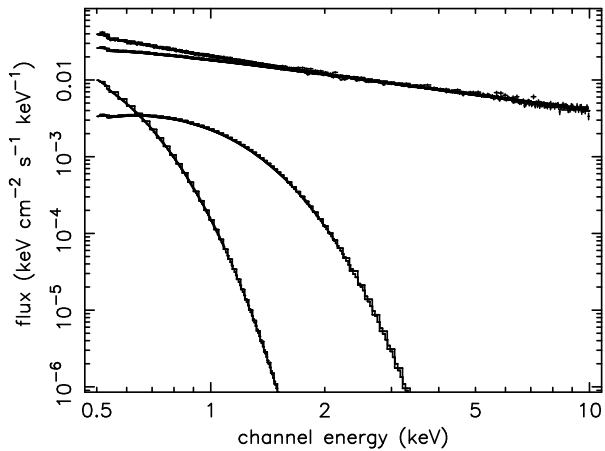}
\caption{An unfolded plot of the blackbody fit to the revolution 95 data. The spectrum is fitted by a power-law and two blackbody components, with kT $\sim$~91 and 232~eV (see Table~\ref{bb}).}
\label{bb_euf}
\end{center}
\end{figure}

Table~\ref{bb} gives the 3--10~keV power-law slopes, together with broad-band BB fits; Fig.~\ref{bb_euf} shows an example of the unfolded BB fit. During revolution 277, observations with both the thin and medium MOS filters were performed, while the same was done for PN during revolution 563; the spectra from the different filters were fitted simultaneously, using the appropriate filter responses, leading to the increased number of degrees of freedom seen in the table. The last column of the table lists the unabsorbed luminosities for the models given. These values show 3C~273 to have been at a typical luminosity during each of the observations, c.f. Courvoisier \etal (1987).

\begin{table*}
\begin{center}
\caption{Power-law and blackbody fits to the MOS 2 and PN 3C~273 spectra. The unabsorbed luminosities are also given in the final column, for the corresponding bandpasses.} 
\label{bb}

\begin{tabular}{p{0.9truecm}p{1.4truecm}p{0.9truecm}p{1.9truecm}p{1.8truecm}p{1.8truecm}p{1.8truecm}p{1.3truecm}p{1.9truecm}}
\hline
Range  & Model & Rev. &  $\Gamma$ & kT (keV) & kT (keV) & kT
(keV) & $\chi^{2}$/dof & luminosity\\

(keV)  & & & & & & & & (10$^{46}$ erg s$^{-1}$)\\
\hline
3--10 & PL & 94 & 1.679~$\pm$~0.007 & & & & 1729/1583 & 0.79\\
 & & 95  & 1.674~$\pm$~0.008 & & & & 1581/1543 & 0.77 \\
 & & 96  &1.643~$\pm$~0.008 & & & & 1685/1548 & 0.76\\
 & & 277 &  1.643~$\pm$~0.006 & & & & 2364/1946 & 0.95\\
 & & 370 & 1.675~$\pm$~0.021 & & & & 901/840 & 1.08\\
 & & 373 & 1.696~$\pm$~0.021 & & & & 818/839 & 1.08\\
 & & 472 & 1.772~$\pm$~0.035 & & & & 464/389 & 0.74\\
 & & 554 &  1.776~$\pm$~0.022 & & & & 873/817 & 1.00\\
 & & 563 & 1.771~$\pm$~0.014 & & & & 1469/1525 & 0.81\\
\hline
0.5--10 & PL + BBs & 94 & 1.679~$\pm$~0.005 & & 0.100~$\pm$~0.003 & 0.252~$\pm$~0.011 & 2802/2331 & 1.68\\
 & & 95 &  1.674~$\pm$~0.020 & & 0.091~$\pm$~0.004 &
0.232~$\pm$~0.010 & 2546/2277 & 1.62\\
 & & 96 &  1.643~$\pm$~0.006 &  & 0.094~$\pm$~0.003 &
0.234~$\pm$~0.012 & 2701/2256 & 1.60\\
 & & 277 & 1.648~$\pm$~0.006  &  0.039~$\pm$~0.002 & 0.117~$\pm$~0.004 &
0.277~$\pm$0.010 & 3833/2741 & 2.06\\
& & 370 & 1.700~$\pm$~0.018 &   & 0.100~$\pm$~0.006 &
0.279~$\pm$~0.028 & 1615/1428 & 2.38\\
& & 373 &  1.699~$\pm$~0.017 & &  0.095~$\pm$~0.007  &
0.261~$\pm$~0.027 & 1621/1426 & 2.33\\
 & & 472 &   1.779~$\pm$~0.035 &  & 0.098~$\pm$~0.010 & 0.330~$\pm$~0.051 & 1134/928 & 1.70\\
  & & 554 &  1.798~$\pm$~0.019 & & 0.098~$\pm$~0.006 & 0.288~$\pm$~0.0024 & 1610/1402 & 2.33\\
 & & 563 & 1.806~$\pm$~0.012 &  & 0.103~$\pm$~0.005 & 0.257~$\pm$~0.016 & 2587/2562 & 1.91\\
\hline
\end{tabular}
\end{center}
\end{table*}

Modelling the soft excess spectrum with multiple blackbodies is probabaly not physical, but it does point to the soft excess being broad. The {\it diskbb} model in {\sc Xspec}
models the accretion disc as emission from multiple blackbody
components, working from the temperature at the inner disc radius
(see, e.g., Mitsuda \etal 1984; Makishima \etal 1986). However, this
method does not succeed in modelling the entire breadth of the
emission seen, and was worse  than either multiple BB (F~=~77, if BB are substituted; $\Delta\chi^{2}$/dof~=~171/2) or the
Comptonisation fits (see below).
An alternative method involves modelling the soft excess with either a second, or a broken, power
law. However, both of these models were
found to be significantly worse fits (the F-test gives improvements of 254 and 124 for the multiple blackbodies in comparison to the two power-laws and broken power-law respectively, for revolution 95 data; these correspond to $\Delta\chi^{2}$/dof of 569/2 and 277/2 respectively), implying that the soft excess
does, indeed, show curvature, rather than a sharp change in slope.

As mentioned above, although multiple blackbodies parametrize AGN soft excesses
very well, the model is not particularly physical. The temperatures
for the hotter components are thought to be far higher than can be
formed through thermal emission from an accretion disc surrounding a
$\sim$~10$^{9}\msun$ black hole (see, e.g., Liu \etal 2003). A more realistic model for the emission is likely to
be Comptonisation: thermal photons, emitted from the disc, are
upscattered by populations of hot electrons. A two-temperature
distribution could then lead to the formation of both the apparent
`power-law' at higher energies and the soft
excess. Low temperature Comptonised spectra are similar in shape to blackbodies, but
are broader; thus, an excess which requires two or three different
temperature BBs can easily be modelled by a single Comptonisation
component. 

To investigate the likelihood of the 3C~273 spectra being
formed via this method, the {\it thCompfe} Comptonisation model
({\.Z}ycki, Done \& Smith 1999) was utilised. Blaes \etal (2001) fit an accretion disc model to multi-wavelength 3C~273 data, finding an accretion rate of 4$\msun$ yr$^{-1}$. This corresponds to a mean disc temperature of 10~eV, which has been used for the X-ray fits presented here, although, in fact, the Comptonisation model is not very sensitive to this value.

If the disc photons are at this representative temperature, then one can consider what would be seen if there were no Comptonising corona and only the direct thermal emission were observed. Taking the disc emission to be characterised by a BB, the resulting peak flux is found to be lower than that predicted by the model in Blaes \etal (2001). That is, more than enough disc photons would be emitted at $\sim$~10~eV to account for the soft excess spectrum observed, showing that our X-ray Comptonisation fits are consistent with the Blaes \etal accretion disc model.

The temperature chosen for the seed photons will affect the integrated flux of the soft excess fitted with this model, but will not affect the values of the other parameters (see, for example, Figure~\ref{OM_BB_comp}). There is no way to determine the temperature from the present data and it has been chosen to fix the value at 10~eV; this is consistent with the OM observations (Figure~\ref{OM_BB_comp}). The absolute level of the soft excess flux is, therefore, somewhat artificial, but the changes in flux between observations will be determined by the measured spectral parameters only.

The electron
temperature of the hotter distribution was fixed at 200~keV. (The
electron temperatures are determined, through spectral fitting, from the energy of the
`roll-off' of the Comptonised component. While it is, therefore, possible to determine an electron temperature for the soft excess, it is expected that the
hard-power-law-producing electrons have very high temperatures, leading to
the $\sim$~4kT roll-off being well outside the {\it XMM} energy band.) Throughout this paper, the {\it hotter} Comptonised component refers to that which forms the power-law observed at the higher energies; the {\it cooler} component produces the soft excess emission.

There are two possible
geometries for the Comptonisation: either (almost) all of the soft
photons are initially Comptonised by the `soft-excess-producing'
electrons; some would then be further Comptonised by a hotter
distribution (possibly formed through magnetic reconnection; these
electrons may be non-thermal), to form the observed `power-law'. An
alternative involves some disc photons being Comptonised to form the
soft excess, while others form the higher
energy spectrum by directly interacting with the hotter electron population. It was found that using the same temperature of input photons to both electron populations did not give as good a fit as having hotter photons pass into the `power-law' electrons. While two Comptonised components have been chosen to model the broad-band spectrum in this analysis, it is, of course, equally possible to represent the high-energy portion as a simple power-law, which might originate as Synchrotron Self-Compton emission in the jet. The general conclusions as to the soft excess reached here are not sensitive to this choice. 

Table~\ref{thcomp} gives the temperatures, optical depths and slopes determined from the Comptonisation fits. Also given is the Compton y-parameter, where y is defined as the average fractional change in energy per Compton scattering multiplied by the mean number of scatterings. From Sunyaev \& Titarchuk (1980), this is given, for an optically thick material of non-relativistic electrons, by

\begin{equation}
y = \frac{4kT}{m_{e}c^{2}} \tau^{2}
\end{equation}

where $\tau$ is the Thomson depth, and kT the temperature, of the electron corona; m$_{e}$ is the mass of an electron. (The optically thin result is the same except the $\tau$ term is not squared.) 

If y~$>>$~1, the Comptonisation process is saturated and results in a Wien-like spectrum ($\sim$~$\nu^{3}$e$^{-\nu}$), with the final temperature of the photons close to that of the electron population. For a low y-parameter, the photons tend to pass straight though the corona, emerging at close to their initial temperature (i.e., that of the accretion disc); this produces a modified BB spectrum. The intermediate regime, where y~$\sim$~1, is known as unsaturated Comptonisation. Here, a power-law spectrum is formed over a limited range; this drops off exponentially for E~$>$~4kT. The resulting spectral index is given by (Sunyaev \& Titarchuk 1980)

\begin{equation}
\Gamma = \left(\frac{4}{y} + \frac{9}{4}\right)^{1/2} - \frac{1}{2}
\end{equation}

\begin{figure}
\begin{center}
\includegraphics[clip,width=0.7\linewidth,angle=-90]{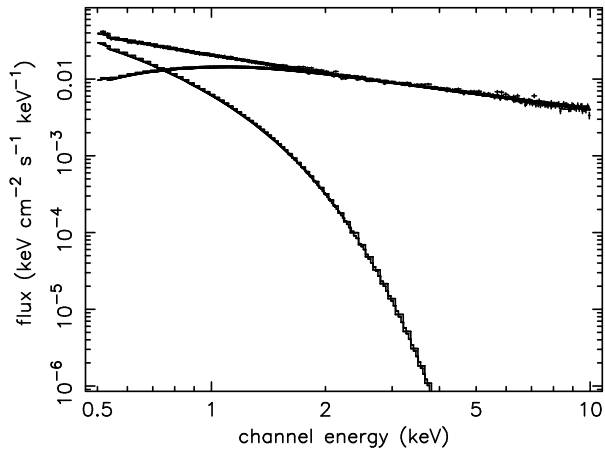}
\caption{The best fit double Comptonisation model to the data from
revolution 95, as given in Table~\ref{thcomp}.}
\label{95_thcomp}
\end{center}
\end{figure}

As Table~\ref{thcomp} shows, the values of y for the soft excess are all $\sim$~0.5, showing the Comptonisation is unsaturated. The corresponding values for the hotter component are slightly larger, $\sim$~1.5, but still in the unsaturated regime. Figure~\ref{95_thcomp} shows the fit to the data from revolution 95.

\begin{table*}
\begin{center}
\caption{Comptonisation model fits. Input BB temperature for the soft excess fixed at 10~eV, that for the hotter component is equal to the `soft excess' temperature; temperature of hotter component is fixed at 200~keV. The luminosities are calculated over the 0.001--10~keV band in the observer's frame, extrapolating the model with the {\it dummyrsp} command.} 
\label{thcomp}
\begin{tabular}{p{0.5truecm}p{2.2truecm}p{1.5truecm}p{1.8truecm}p{1.9truecm}p{1.8truecm}p{1.9truecm}p{2.0truecm}}
\hline
 &   \multicolumn{4}{c} {\sc cooler comptonised component} & \multicolumn{2}{c}{\sc hotter comptonised  component}\\
Rev.  & electron  & $\tau$ & y-parameter  & luminosity & $\Gamma$ &  luminosity & $\chi^{2}$/dof \\
&  temp. (keV) & & & (10$^{46}$ erg s$^{-1}$) & &  (10$^{46}$ erg s$^{-1}$)\\
\hline

94 &  0.292~$\pm$~0.014 & 15.1~$\pm$~0.6 & 0.52~$\pm$~0.01 & 4.34~$\pm$~0.37 & 1.693~$\pm$~0.005 & 1.52~$\pm$~0.06 & 2805/2332\\
95 &  0.289~$\pm$~0.013 & 15.4~$\pm$~0.5 & 0.54~$\pm$~0.01 & 3.92~$\pm$~0.33 & 1.683~$\pm$~0.005 &  1.48~$\pm$~0.05 & 2556/2278\\
96 &  0.296~$\pm$~0.017 & 14.5~$\pm$~0.6 & 0.49~$\pm$~0.01 & 4.93~$\pm$~0.50 & 1.663~$\pm$~0.005 &  1.45~$\pm$~0.06 & 2703/2257\\
277 & 0.334~$\pm$~0.011 & 14.2~$\pm$~0.3 & 0.53~$\pm$~0.01 & 6.35~$\pm$~0.28 & 1.667~$\pm$~0.005 & 1.75~$\pm$~0.06 & 3856/2844\\
370 &  0.399~$\pm$~0.040 & 13.5~$\pm$~1.1 & 0.57~$\pm$~0.03 & 5.81~$\pm$~0.70 & 1.697~$\pm$~0.018 & 1.93~$\pm$~0.23 & 1611/1429\\
373 &  0.353~$\pm$~0.036 & 14.1~$\pm$~1.1 & 0.55~$\pm$~0.03 & 6.50~$\pm$~1.14 &1.704~$\pm$~0.016 & 2.05~$\pm$~0.19 & 1624/1427\\
472 & 0.422~$\pm$~0.072 & 13.3~$\pm$~1.8 & 0.59~$\pm$~0.05 & 4.16~$\pm$~0.87 &1.782~$\pm$~0.031 &  1.34~$\pm$~0.32 & 1134/929\\
554 &  0.400~$\pm$~0.036 & 13.5~$\pm$~0.9 & 0.57~$\pm$~0.02 & 6.59~$\pm$~0.69 &1.793~$\pm$~0.018 & 1.84~$\pm$~0.21 & 1610/1403\\
563 &  0.331~$\pm$~0.020 & 14.4~$\pm$~0.7 & 0.54~$\pm$~0.01& 6.35~$\pm$~0.57 & 1.803~$\pm$~0.011 & 1.58~$\pm$~0.11 & 2585/2563\\

\hline
\end{tabular}
\end{center}
\end{table*}

\begin{figure}
\begin{center}
\includegraphics[clip,width=0.7\linewidth,angle=-90]{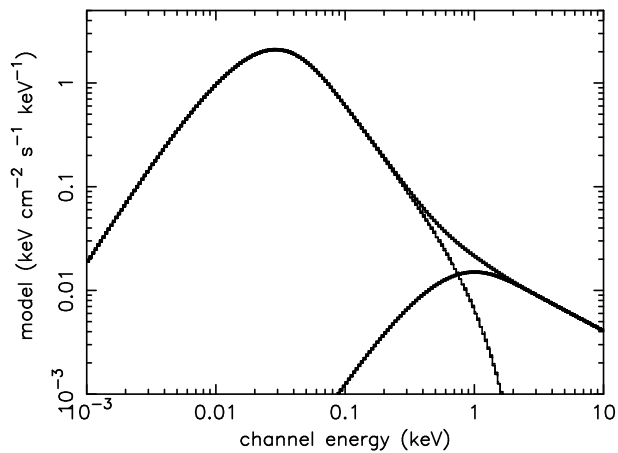}
\caption{Extrapolating the revolution 95 {\it thComp} fit down to lower energies shows the full soft excess. The lowest energy is fixed by the temperature set for the input photons, which cannot be determined over the {\it XMM} energy band.}
\label{95_model}
\end{center}
\end{figure}

Although data have only been modelled down to 0.5~keV in this analysis, the Comptonisation model used here covers a much broader energy band, down to the 10~eV of the seed photons from the accretion disc. Figure~\ref{95_thcomp} shows that the soft Comptonisation component is still rising at 0.5~keV, so its luminosity over the {\it XMM} band will be much less than the total value. In order to obtain an estimate for the total luminosity of the soft excess, the model was extrapolated to lower energies, using the {\it dummyrsp} command in {\sc Xspec}; an example of such a model plot is shown in Figure~\ref{95_model}. Note that, as discussed earlier, 10~eV is not a fitted parameter, but has simply been chosen to represent a suitable temperature for the disc, not inconsistent with the data obtained from the Optical Monitor (Section~\ref{sec:om}). This extrapolation to lower temperatures does not affect the hotter Comptonised component greatly. Table~\ref{flux} lists the fluxes over the 3--10, 0.5--10 and 0.001--10~keV bands for each observation.

\begin{table}
\begin{center}
\caption{Observed fluxes calculated over the 0.5--10 and 0.001--10~keV (both observer's frame) energy bands.} 
\label{flux}
\begin{tabular}{p{1.0truecm}p{2.5truecm}p{2.5truecm}}
\hline
Rev.  & 0.5--10~keV flux & 0.001--10~keV flux\\
 & \multicolumn{2}{c}{(photon cm$^{-2}$ s$^{-1}$)}\\ 
\hline
94 &  0.013~$\pm$~0.001 & 5.074~$\pm$~0.429\\
95 &  0.014~$\pm$~0.001 & 4.516~$\pm$~0.382\\
96 &  0.012~$\pm$~0.001 & 6.014~$\pm$~0.610\\
277 &  0.022~$\pm$~0.001 & 7.325~$\pm$~0.326\\
370 &  0.027~$\pm$~0.003 & 6.219~$\pm$~0.747\\
373 &  0.024~$\pm$~0.004 & 7.550~$\pm$~1.325\\
472 & 0.022~$\pm$~0.005 & 4.347~$\pm$~0.913\\
554 &  0.031~$\pm$~0.003 & 7.062~$\pm$~0.736\\
563 &  0.023~$\pm$~0.002 & 7.238~$\pm$~0.649\\
\hline
\end{tabular}
\end{center}
\end{table}

\section{RGS data}
\label{sec:rgs}

To a first approximation, the soft excess of 3C~273 can be modelled as a smooth continuum in the {\it XMM} band. However, it has previously been found that some AGN show features in their soft excesses: either (possible) relativistic emission lines (Branduardi-Raymont \etal 2001; the existence of these lines remains controversial, with Lee \etal 2001 claiming the spectrum can be explained with a dusty warm absorber) or a combination of narrow emission/absorption features, sometimes with an absorption trough around 16--17~\AA (Sako \etal 2001; O'Brien \etal 2001; Pounds \etal 2001; Kaspi \etal 2000). 

The 3C~273 RGS (den Herder \etal 2001) spectrum during revolution 277 was chosen for analysis, since this had the longest duration, of almost 90~ks. 3C~273 clearly does not exhibit the same spectral shape as that found by Branduardi-Raymont \etal (2001) for MCG~$-$6$-$30$-$15 and Mrk~766. To investigate whether any  narrow features could be found, the lines identified in Mrk~359 (O'Brien \etal 2001) were considered. Weak, but statistically significant, emission from the triplets of Ne~{\sc ix} and O~{\sc vii} was identified in the RGS spectrum of 3C~273. As for Mrk~359, the individual components cannot be resolved; however, their combined equivalent widths are (3.5~$\pm$~0.9)~eV and (1.6~$\pm$~0.6)~eV respectively. Only upper limits were obtained for the other features found in Mrk~359, with O~{\sc viii}~Ly$\alpha$ having an equivalent width of EW~$<$~0.06~eV (90~per~cent upper limit). There is no sign of an Fe~M absorption trough (EW~$<$~0.61~eV) or absorption edges corresponding to O~{\sc vii} or O~{\sc viii}. The soft excess is, therefore, not dominated by a blend of soft X-ray lines (Turner \etal 1991). If, as is expected, the soft excess continuum is formed through Comptonisation, it would be expected that any emission lines would be greatly broadened (Sunyaev \& Titarchuk 1980), which would explain the lack of strong emission observed in most AGN. 

\begin{figure}
\begin{center}
\includegraphics[clip,width=0.7\linewidth,angle=-90]{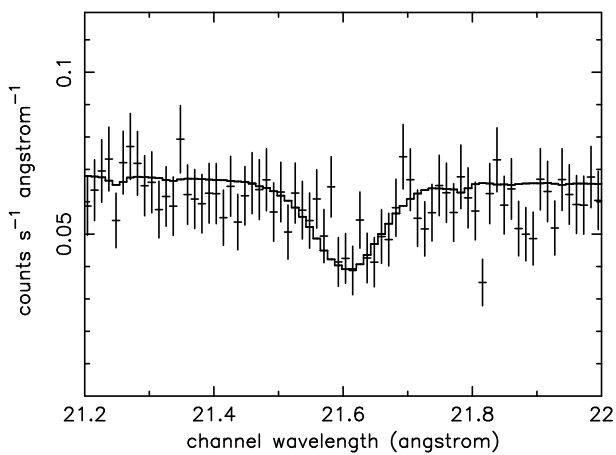}
\caption{The plot shows an absorption feature in the RGS data which may be due to O~{\sc vii} He$\alpha$ at zero redshift. The data shown are from the revolution 277 observation}
\label{3c_rgs}
\end{center}
\end{figure}

Fang, Sembach \& Canizares (2003) claim to find an absorption feature, in {\it Chandra} data, corresponding to the zero-redshift O~{\sc vii} He$\alpha$ ($\sim$~21.60~\AA), for which they give an equivalent width of $\sim$~28~m\AA ($\Leftrightarrow$~0.75~eV). As Fang \etal (2003) discuss, this is likely to correspond to the detection of warm gas in the local intergalactic medium. Investigating the possibility of absorption in the revolution 277 {\it XMM} data, a feature is, indeed, found: at a wavelength of 21.62~$\pm$~0.16, with a line width of 1~eV, the equivalent width is $\sim$0.88~eV -- similar to the value in Fang \etal (2003). This line is detected at the $>$99.99~per~cent level and is shown in Figure~\ref{3c_rgs}. Similar zero-redshift absorption is also discussed by Rasmussen, Kahn \& Paerels (2003), who analyse both {\it Chandra} and {\it XMM} grating data.

\section{Optical and UV data}
\label{sec:om}

For a number of the observations (Revolutions 94--277, 382 and 563), optical/UV data were obtained from the Optical Monitor (OM; Mason \etal 2001); the remaining orbits had the OM in Grism mode. The observed magnitudes for the various filters used are listed in Table~\ref{om}.

\begin{table*}
\begin{center}
\caption{Optical and UV magnitudes for 3C~273.} 
\label{om}
\begin{tabular}{p{1.5truecm}p{2.0truecm}p{1.0truecm}p{1.0truecm}p{1.0truecm}p{1.2truecm}p{1.2truecm}p{1.2truecm}}
\hline
& &  Rev. 94 & Rev. 95 & Rev. 96 & Rev. 277 & Rev. 382 & Rev. 563\\ 
  Filter & Wavelength ({\AA})& \multicolumn{6}{c} {Magnitude}\\

\hline
V & 5500 & 12.54 & 12.60 & 12.58 & 12.26 & 12.67 & -\\
U & 3600 & 11.90 & 11.94 & 11.97 & 11.66 & - & -\\
B & 4400 & 13.24 & 13.01 & 12.98 & 13.10 & - & -\\
UVW1 & 2910 & 11.78 & 12.09 & 11.81 & 11.03 & - & 11.36\\
UVM2 & 2310 & 11.75 & 11.81 & 11.77 & 11.16 & - & 11.21\\
UVW2 & 2120 & 11.72 & 11.73 & 11.74 & 11.11 & - & -\\
\hline
 \end{tabular}
\end{center}
\end{table*}

\begin{figure}
\begin{center}
\includegraphics[clip,width=0.7\linewidth,angle=-90]{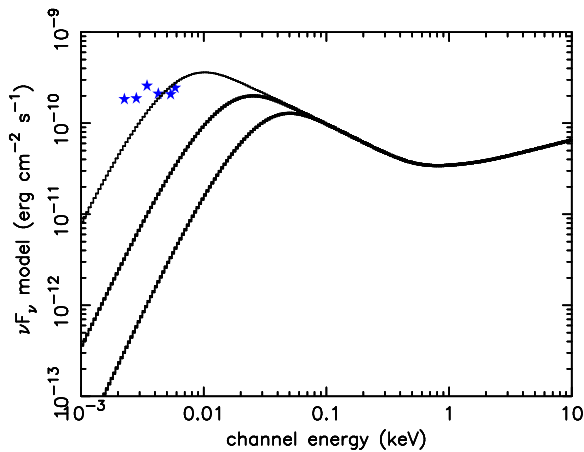}
\caption{The curves show the extrapolation of the X-ray Comptonisation model down to optical energies for input (disc) temperatures of, from left to right, 2, 5 and 10~eV. The data are taken from revolution 95.}
\label{OM_BB_comp}
\end{center}
\end{figure}

A lower limit to the temperature of the accretion disc can be roughly estimated by finding the point at which the extension of the X-ray fit would produce a higher optical flux than is observed by the OM. Figure~\ref{OM_BB_comp} plots the extrapolated X-ray Comptonisation model for three different input temperatures (10, 5 and 2~eV). For temperatures below $\sim$2~eV, the model over-predicts the optical flux. This can, therefore, be taken as a lower limit to the temperature of the photons producing the soft excess.

%The top plot assumes the input photons follow a simple blackbody, while the lower one has a disc blackbody as the input to the model.

However, it must be noted that 2~eV is far too low a temperature for an accretion disc such as this. Disc theory is still a very uncertain area, especially at high accretion rates. Collin \& Hur{\'e} (2001) and Collin \etal (2002) find that the optical emission in AGN cannot be accounted for by the standard accretion disc model, concluding that, either the disc is  `non-standard' or most of the optical luminosity does not come from the accretion disc. A much improved theory of accretion discs is clearly needed, before the relationship between optical and X-ray emission in AGN can be fully understood.

\section{Discussion}
\label{sec:disc}

\subsection{3--10~keV band}

As mentioned in the introduction (Section~\ref{sec:intro}), 3C~273 has, in the past, shown a hard spectral index of $\sim$~1.5 (or flatter), while the values found here (Table~\ref{bb}) are noticeably steeper. This is not likely to be due to contamination by the broad soft excess, since the values of $\Gamma$ for the broad-band power-law plus BB fits are very similar to the 3--10~keV slope. Neither does it appear to be a calibration problem, since Molendi \& Sembay (2003) also find a relatively steep slope ($\Gamma$~$\sim$~1.63) for the MECS spectrum over 3--10~keV ({\it SAX} data simultaneous with {\it XMM} revolution 277).
 
\begin{figure}
\begin{center}
\includegraphics[clip,width=0.7\linewidth,angle=-90]{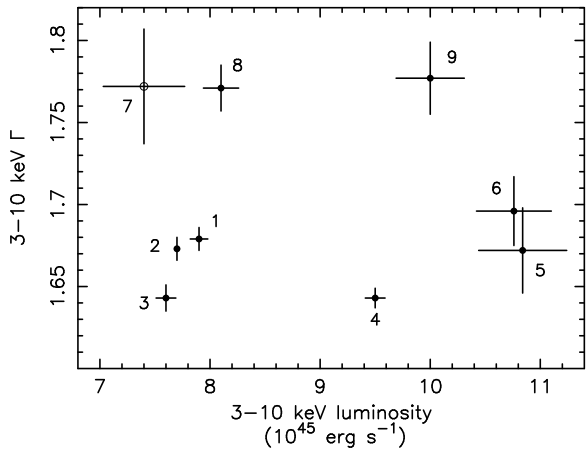}
\caption{If all nine observations are considered, the 3--10~keV photon index is independent of the luminosity over that range. There is a suggestion, however, that 3C~273 exists in two different `hardness states'. The open circle represents the revolution 472 data point, while the numbers indicate the chronological order of the measurements.}
\label{3-10gamma-lum}
\end{center}
\end{figure}

Using the simple power-law model (Table~\ref{bb}), there is no strong correlation between 3--10~keV $\Gamma$ and the power-law luminosity
over the same energy band (Figure~\ref{3-10gamma-lum}), with Spearman Rank giving a negligible probability of 9~per~cent. However, it is noticeable that the first six observations cluster around a 3--10~keV slope of $\Gamma$~$\sim$~1.66, while the later three show steeper values, of $\sim$~1.77.

The open circle in this, and subsequent, plots denotes the data from revolution 472. As mentioned earlier, of the nine observations presented here, it is during revolution 472 that 3C~273 is at its faintest over the 3--10~keV band; revolution 472 is also the shortest of all the observations. Ignoring this data point does not, however, lead to a significant correlation between the 3--10~keV photon index and flux (57~per~cent). 
The relationship between the slope and flux of 3C~273 has been found to vary, however. Agreeing with the present data Turner \etal (1990) found that the parameters were independent, when considering {\it Ginga} data between 1983 and 1988; this was also found from {\it RXTE} measurements from 1996-1997 (Kataoka \etal 2002). However, from 1999-2000, Kataoka \etal found that the slope became {\it softer} when the flux level increased.

\subsection{Soft excess}
\subsubsection{0.5-10 keV band}

\begin{figure}
\begin{center}
\includegraphics[clip,width=0.7\linewidth,angle=-90]{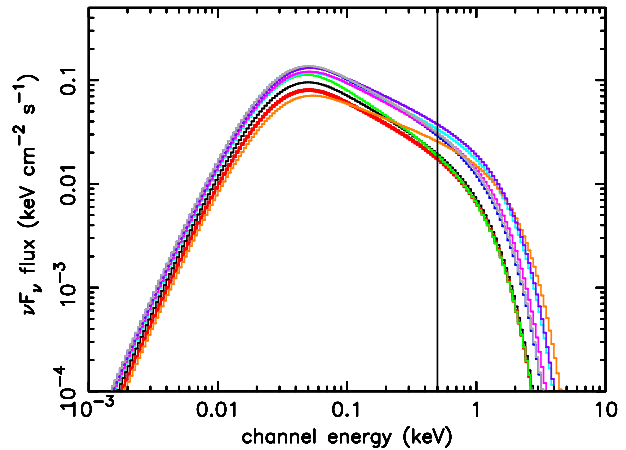}
\caption{The soft excess of 3C 273 measured during each observation. The vertical line at 0.5 keV shows the point above which the model was fitted; below this energy, the curves are purely an extrapolation. Each revolution is shown in a different colour: black -- rev. 94; red -- rev. 95; green -- rev. 96; blue -- rev. 277; light blue -- rev. 370; magenta -- rev. 373; orange -- rev. 472; purple -- rev. 554; grey -- rev. 563}
\label{eemodel}
\end{center}
\end{figure}

To investigate how the soft excess is changing over time, the Comptonisation model plots for each revolution were overlaid; the result is shown in Figure~\ref{eemodel}. Above 0.5~keV the fitted spectra show a general tendency for higher flux to be associated with higher temperatures -- the hotter the soft excess, the higher the energy the model extends to. Revolution 472 (orange curve) is the exception here, with a high temperature, but a lower normalisation. However, when the model is extrapolated to lower energies, the different $\Gamma$s become more important, the major contribution to the model luminosity being below 0.5~keV and sensitive to the value of $\Gamma$, rather than to kT.

\begin{figure}
\begin{center}
\includegraphics[clip,width=0.7\linewidth,angle=-90]{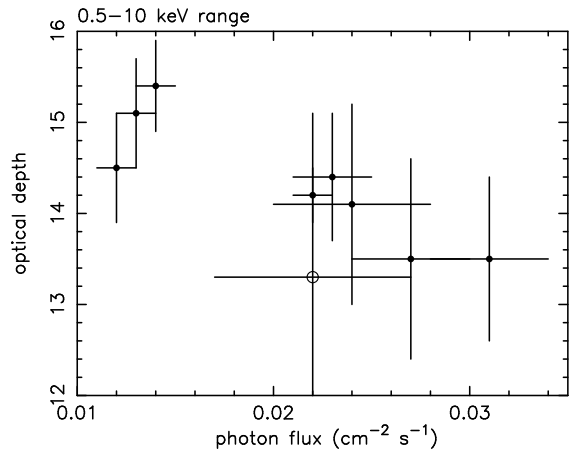}
\caption{There is an apparent negative correlation between the optical depth and the flux of the Comptonised soft excess parameter. The open circle represents the revolution 472 data point, as explained in the text.}
\label{tau}
\end{center}
\end{figure}

\begin{figure}
\begin{center}
\includegraphics[clip,width=0.7\linewidth,angle=-90]{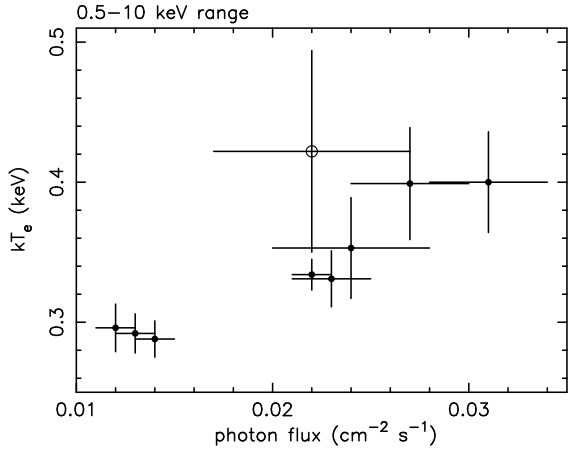}
\caption{Over the observed 0.5--10~keV band, when the soft excess is brighter, it is also hotter. The open circle represents the revolution 472 data point.}
\label{kT}
\end{center}
\end{figure}

In order to investigate how the soft excess varies, the parameters ($\Gamma$, optical depth and electron temperature) were plotted against the photon flux calculated over the 0.5--10~keV (observer's frame) band. Figures~\ref{gamma}, \ref{tau} and \ref{kT} show these results. Spearman Rank (SR) and weighted linear regression were then used to give the probability of a correlation. Linear regression is a useful method, since it takes into account the errors on the measurements, whereas Spearman Rank does not.

%Plotting the soft excess $\Gamma$ against its flux over 0.5--10~keV (Figure~\ref{gamma}), a negative correlation is found (97~per~cent probability from Spearman Rank; linear regression gives a slope of $-$4.9~$\pm$~2.0). If the data from revolution 472 are excluded, this correlation becomes stronger (99.9~per~cent; the regression slope stays approximately the same, since the point being ignored has relatively large error bars, so would have been less strongly weighted than the other measurements).

 An inverse relation is found between the optical depth and the flux over 0.5--10~keV (98~per~cent; slope of $-$88~$\pm$~38) and is shown in Figure~\ref{tau}. Again, excluding the revolution 472 point strengthens the Spearman Rank value, to 99.8~per~cent. (The regression slope stays approximately the same, since the point being ignored has relatively large error bars, so would have been less strongly weighted than the other measurements)  The other physical parameter of the soft excess -- the temperature -- shows a 98~per~cent positive correlation with the flux (99.8~per~cent without revolution 472); that is, the hotter soft excesses are brighter. This is supported by a regression fit of 5.5~$\pm$~1.3. It must, however, be cautioned that the temperature, kT, and the optical depth, $\tau$, are strongly coupled, as shown by Equations 1 and 2: $\Gamma$~$\sim$~1/(kT$\tau^2$)$^{1/2}$.

%\begin{figure}
%\begin{center}
%\includegraphics[clip,width=0.7\linewidth,angle=-90]{softexcessgamma_vs_0.5-10photons_jointfit.ps}
%\caption{There is an inverse correlation between the slope of the soft excess and its flux over 0.5--10~keV: the flatter soft excesses are brighter in this band. The open circle represents the revolution 472 data point, as explained in the text.}
%\label{gamma}
%\end{center}
%\end{figure}

\begin{figure}
\begin{center}
\includegraphics[clip,width=0.7\linewidth,angle=-90]{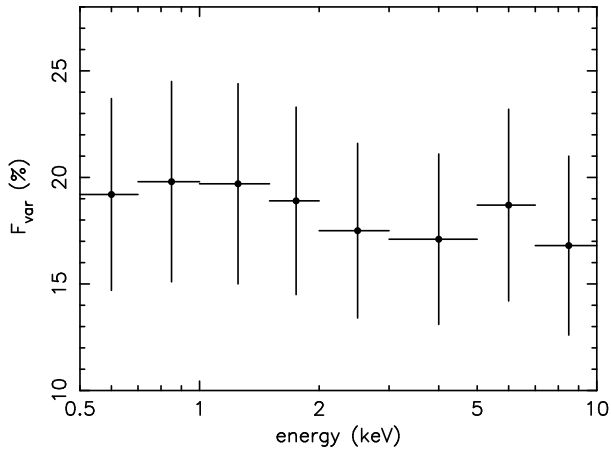}
\caption{The fractional variabilty amplitude of the MOS 2 data over the nine observations of 3C~273 showing that there is no difference in the variability observed over the different energies. The PN data are consistent with these results.}
\label{rms}
\end{center}
\end{figure}

To try and determine how the regions producing the soft and hard
X-rays are linked, Figure~\ref{rms} was plotted. This shows the fractional (rms) variability of the total observed counts as a function of energy, after the Poisson noise has been accounted for (Edelson \etal 2002) for the MOS data and implies that, over the nine observations, there is very little difference between the rate of variation in the soft and hard bands. The PN results also show that rms variability does not change with energy, although the fraction variability is higher ($\sim$~32~per~cent, compared to $\sim$~18~per~cent in the MOS 2 datasets). 

\begin{figure}
\begin{center}
\includegraphics[clip,width=0.7\linewidth,angle=-90]{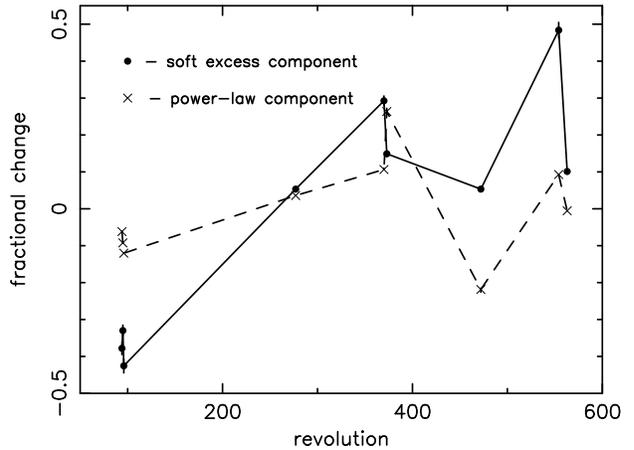}
\caption{Considering the flux of the soft and hard {\it thComp} components, the fractional change between revolutions was plotted. In this figure, the solid line joins the soft excess points, while the dashed line joins the `power-law' component values. There is a larger variability in the soft flux, than the hard. The errors on the values are all very small ($<$0.02).}
\label{frac}
\end{center}
\end{figure}

Figure~\ref{frac} shows how the two separate Comptonised components vary in flux during the nine observations. It can be seen that the soft excess component varies over a larger range than does the hotter component, although the two generally change in the same sense. This is shown by the standard deviation values of $\sigma$~$\sim$~0.3 for the range in soft excess variation, compared to $\sigma$~$\sim$~0.1 for the hotter, power-law component.

\subsubsection{0.001--10~keV band}

%Figure~\ref{extrapgamma} shows that the strong negative correlation between the soft excess $\Gamma$ and the photon flux over 0.5--10~keV disappears when considering the integrated flux over the 0.001--10~keV band: there is only a 27~per~cent probability of a {\it positive} correlation.

Cutting the spectrum at 0.5~keV could give a wrong impression of the total luminosity of the soft excess component in the Comptonisation model, since most of the emitted flux is below this energy. For this reason, the optical depth and temperature were plotted against the flux of the extrapolated soft Comptonised component, following the method for investigating the 0.5--10~keV energy band. It must be cautioned, however, that the fluxes and luminosities derived for this extended band are very sensitive to the parameters of the fit.

\begin{figure}
\begin{center}
\includegraphics[clip,width=0.7\linewidth,angle=-90]{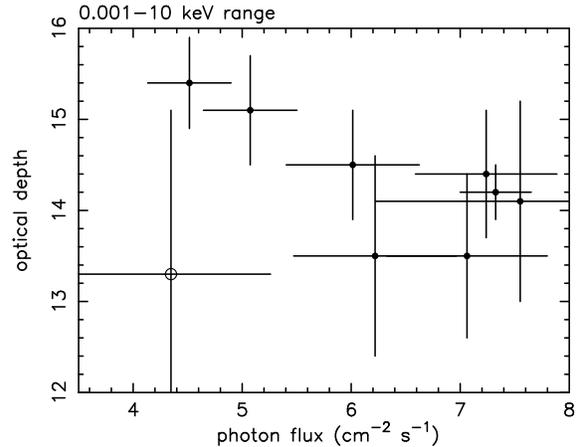}
\caption{The open circle represents the revolution 472 data point. If this is excluded, then there is a weak anti-correlation between the optical depth and the photon flux over 0.001--10~keV.}
\label{extraptau}
\end{center}
\end{figure}

\begin{figure}
\begin{center}
\includegraphics[clip,width=0.7\linewidth,angle=-90]{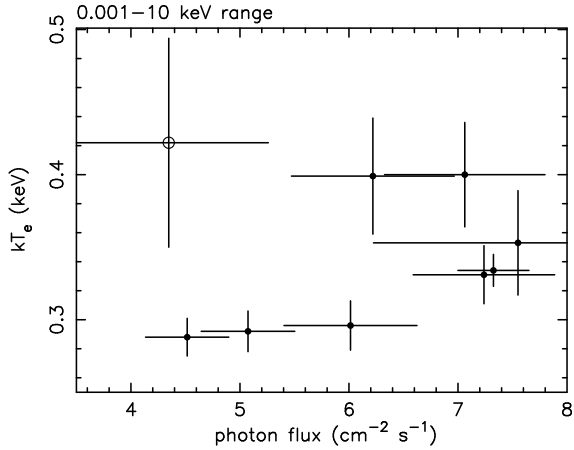}
\caption{There may be a small, positive correlation between the temperature of the soft excess and the extrapolated photon flux. The open circle represents the revolution 472 data point.}
\label{extrapkt}
\end{center}
\end{figure}

The correlation between the optical depth and the photon flux (Figure~\ref{extraptau}) becomes much weaker, with the Spearman Rank probability being 30~per~cent for an inverse relation between the two parameters. The kT-flux is also only present at the 30~per~cent level, as shown in Figure~\ref{extrapkt}. Neither of these probabilities is significant. If, however, the revolution 472 data point is, again, excluded from the calculation, weak correlations between the photon flux and soft excess temperature or optical depth are revealed (93~per~cent, positive/negative for kT/$\tau$ respectively; using linear regression, best-fit slopes of 0.021~$\pm$~0.006 and $-$0.43~$\pm$~0.19 are obtained). These correlations are weaker over the full energy band, but are in the same sense as the 0.5--10~keV band. 
%For $\Gamma$, the relatively strong correlation between it and the 0.5--10~keV flux entirely disappears when considering the 0.001--10~keV range.

%\begin{figure}
%\begin{center}
%\includegraphics[clip,width=0.7\linewidth,angle=-90]{softexcessgamma_vs_0.001-10photons_jointfit.ps}
%\caption{When considering the flux calculated for the extrapolated model, there is no longer a correlation between $\Gamma$ and photon flux. The open circle represents the revolution 472 data point.}
%\label{extrapgamma}
%\end{center}
%\end{figure}

\begin{figure}
\begin{center}
\includegraphics[clip,width=0.7\linewidth,angle=-90]{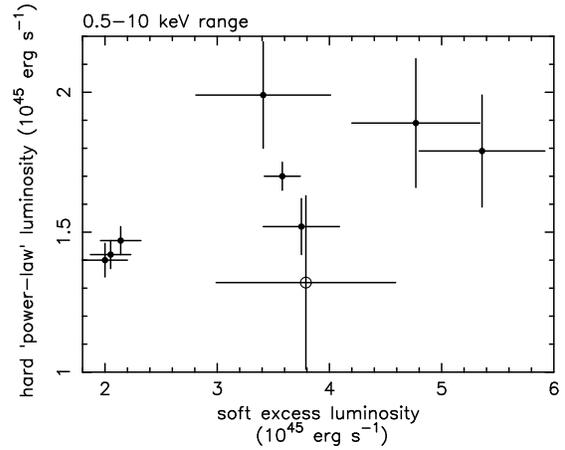}\vspace*{0.5cm}
\includegraphics[clip,width=0.7\linewidth,angle=-90]{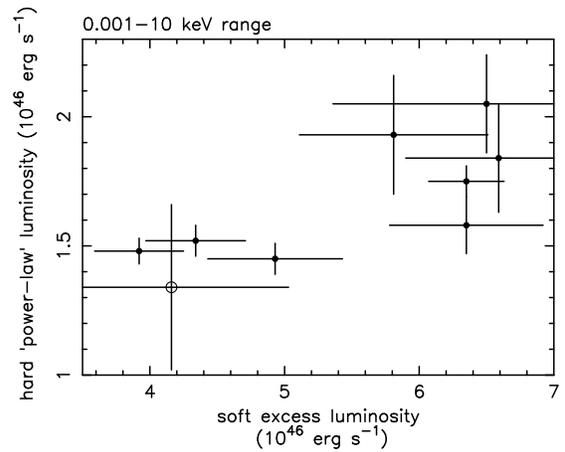}
\caption{There is a positive correlation between the individual Comptonised components, with least squares fitting giving similar slopes of 0.15~$\pm$~0.03 and 0.13~$\pm$~0.03 for the 0.5--10 and 0.001--10~keV ranges respectively. The open circle represents the revolution 472 data point.}
\label{lumvariation}
\end{center}
\end{figure}

\begin{figure}
\begin{center}
\includegraphics[clip,width=0.7\linewidth,angle=-90]{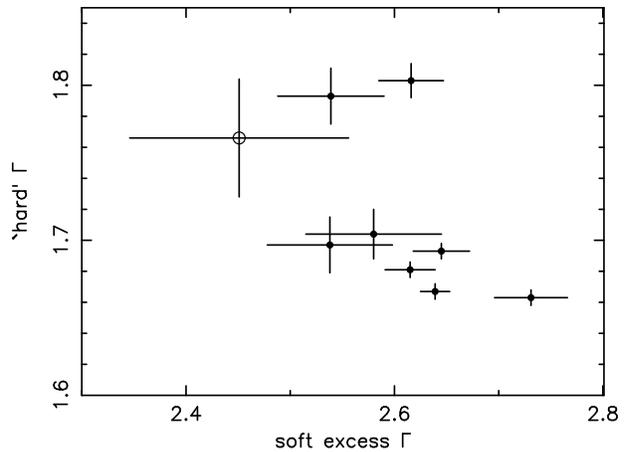}
\caption{When fitting the spectra with two Comptonisation components, there is a negative correlation between the values of $\Gamma$: $-$1.64~$\pm$~0.33. The open circle represents the revolution 472 data point.}
\label{gammavariation}
\end{center}
\end{figure}

Figures~\ref{lumvariation} and \ref{gammavariation} plot the luminosities and slopes of the two Comptonised components. The soft/hard luminosities appear to be correlated over both bands if weighted linear regression is used to give the line of best fit ($\sim$~0.14 over both bands); Spearman Rank, however, gives an insignificant result of 74~per~cent for the 0.5--10~keV band, though this increases to 96~per~cent if the revolution 472 point is ignored. Over 0.001--10~keV, Spearman Rank gives a much larger probability of 99~per~cent for a positive correlation (approximately constant with or without revolution 472). The slopes of the cooler and hotter Comptonised components appear to be inversely related, with linear regression giving a slope of $\sim$~$-$1.64, and Spearman Rank giving a (fairly low) probability of 92~per~cent for a negative correlation.

\subsection{Compton cooling}

It is natural to suppose that, if the soft excess were to be produced by Comptonisation of thermal photons from the accretion disc, then an inverse relationship would exist between the photon flux and the electron temperature, caused by Compton cooling. The positive correlation found between kT and the flux over 0.5--10~keV for the soft excess could well have been an artifact of the 0.5~keV cut-off. However, the same sense is observed in the correlation over the full energy band. Thus, the data indicate no negative relationship between the flux and kT and, therefore, no Compton cooling. This implies that the soft excess, if produced by the Comptonisation of thermal disc photons, is more complex than the simple model proposed here. The total soft excess varies from 3.3~$\times$~10$^{46}$ erg~s$^{-1}$ up to 6.6~$\times$~10$^{46}$ erg~s$^{-1}$, but this is accompanied by a modest increase in temperature. This implies both an increase in the thermal disc emission and a corresponding {\it increase} in the electron temperature: two separate mechanisms need to be invoked. Note that the data are inconsistent with a constant {\it photon flux} but varying temperature (Figure~\ref{extrapkt}).

\section{Conclusions}
\label{sec:conc}

The {\it XMM} spectrum of 3C~273 has been investigated. It is found that the soft X-ray spectrum is dominated by a strong soft excess below $\sim$~2~keV. This can be well modelled by a multiple blackbody parametrisation, but is most likely to arise though thermal Comptonisation of cool (UV) disc photons in a warm (few hundred eV) corona above the surface of the accretion disc. While the soft excess spectra can be fitted with the Comptonisation model, the variability behaviour is not consistent with a simple interpretation of this model. The temperature of the Comptonising electron cloud may vary independently of the input photon flux, or may even be positively correlated with it. If the latter were to be true, then a further link between the disc emission and the energising of the Comptonising electrons would be necessary.

The individual spectra do not tend to show iron lines, either narrow or broad, neutral or ionised. However, if all the observations are co-added, a weak, broad
line is detected.

\section{ACKNOWLEDGMENTS}
The work in this paper is based on observations with {\it
XMM-Newton}, an ESA
science mission, with instruments and contributions directly funded by
ESA and NASA. The authors would like to thank the EPIC Consortium for all their work during the calibration phase, 
and the SOC and SSC teams for making the observation and analysis
possible. 
This research has made use of the NASA/IPAC Extragalactic
Database (NED), which is operated by the Jet Propulsion Laboratory,
California Institute of Technology, under contract with the National
Aeronautics and Space Administation.
Support from a PPARC studentship is gratefully acknowledged by KLP.

\end{document}